\documentclass[aps,pre,twocolumn,superscriptaddress]{revtex4}
\usepackage{graphicx,amssymb,amsmath,color,float,mathtools}
\usepackage{enumitem}
\usepackage{inputenc}
\usepackage{blindtext}
\usepackage{booktabs,multirow}
\usepackage{color,soul}
\usepackage{amsthm,mathrsfs,amsfonts,dsfont,bm}
\usepackage{tasks}
\usepackage[british]{babel}
\usepackage[title]{appendix}
\usepackage{caption}
\usepackage[export]{adjustbox}
\usepackage{subfig}
\usepackage{xcolor}
\usepackage{hyperref} 

\begin{document}
	
	\title{Mean-Field Solution of Structural Balance Dynamics in Non-Zero Temperature }
	\author{F. Rabbani}
	\email{$fereshte.rabbani@gmail.com$}
	\affiliation{Department of Physics, Shahid Beheshti University, G.C., Evin, Tehran, 19839, Iran}
	\author{Amir H. Shirazi}
	\affiliation{Department of Physics, Shahid Beheshti University, G.C., Evin, Tehran, 19839, Iran}
	\author{G.R. Jafari}
	\affiliation{Department of Physics, Shahid Beheshti University, G.C., Evin, Tehran, 19839, Iran}
	
	\begin{abstract}
			
In signed networks with simultaneous friendly and hostile interactions, there is a general tendency to a global structural balance, based on the dynamical model of links status. Although, the structural balance represents a state of the network with a lack of contentious situations, there are always tensions in real networks.
To study such networks, we generalize the balance dynamics in non-zero temperatures. The presented model uses elements from Boltzmann-Gibbs statistical physics to assign an energy to each type of triad, and it introduces the temperature as a measure of tension tolerance of the network. Based on the mean-field solution of the model, we find out that the model undergoes a first-order phase transition from an imbalanced random state to structural balance with a critical temperature $T_c$, where in the case of $T>T_c$ there is no chance to reach the balanced state. A main feature of the first-order phase transition is the occurrence of a hysteresis loop crossing the balanced and imbalanced regimes.\\
\url{https://doi.org/10.1103/PhysRevE.99.062302}
	\end{abstract}

	\maketitle
	
	\section{Introduction}

	Signed social networks with both positive and negative links are used to indicate the relationships between people, such as friendship-like and animosity-dislike. A basic network analysis method to perceive such relations is “structural balance theory” originally proposed by Heider \cite{Heider}. Cartwright and Harary \cite{Cartwright} further developed it in terms of signed graphs. Balance theory has been applied in many fields, such as social, economic, ecologic, and political systems\cite{hart1974symmetry,hummon2003some,szell2010multirelational,facchetti2011computing,lerner2016structural,saiz2017evidence,saeedian2017epidemic}. Balance theory is used to describe attitudes of individuals to reduce tension among each other and measure social balance in a given signed network. In this respect, we provide a further expansion of balance theory utilizing methods from Boltzmann-Gibbs statistical physics to show the remarkable influence of temperature on the evolution of networks.

According to structural balance theory, when people set up dyadic relations that contain both positive and negative interactions, four different types of triads would be created (Fig.(\ref{fig1})).
		\textcolor{black}{Among these four possibilities, certain configurations are more socially and psychologically plausible than others. We distinguish the following cases:}
		
	\begin{itemize}
    	\item 	\textcolor{black}{Three positive relationships draw on a balanced state, or a situation that is psychologically plausible. It represents three people who are mutual friends.}
    	
		\item 	\textcolor{black}{We also have a balanced state when there are two negative relationships with one positive relationship. It means that two of the three are friends, and they have a mutual enemy.} 
		
		\item 	\textcolor{black}{The other two possible triads introduce some amount
		of psychological “tension” or “instability” into the relationships. It shows two people who are enemies but they have a mutual friend. In such cases, there would be implicit forces pushing them to become friends thus turning the negative relationship into a positive, otherwise their mutual friend will side with one of them and against the other (turning one of the positive relationships into a negative).}
	
		\item 	\textcolor{black}{In the same way, there are sources of stress in a configuration where all the people are mutual enemies. In such cases, there would be forces motivating two of the three people to work together against the third one (turning one of the three relationships into a positive one).}
	\end{itemize}
	
        \textcolor{black}{Based on this reasoning, we refer to triads with one or three positive relationships as balanced, since they are free of these sources of stress, and we refer to triads with zero or two positive relationships as imbalanced. The structural balance hypothesis is that since imbalanced
	 	triads are sources of stress and generate tension for the people
	 	involved, they attempt to minimize them in their personal relationships, and hence they will be slighter in real social networks than balanced triads.}
	
	The central notion of balance theory is that a network of signed relations has a tendency towards a more balanced situation \cite{marvel,abell2009structural,leskovec2010signed,traag2013dynamical}. Such a network is considered structurally balanced if either all individuals are friends or if there are two antagonistic cliques, with friendly relations within each clique and all pairs of persons in different cliques being enemies (bipartite) \cite{Antal,davis1967clustering,marvel2011continuous,hedayatifar2017pseudo,e19060246}. Although, the bipartite state happens more often  in real social networks \cite{langer1878european}.
	
	 Early studies of balance theory focused on static properties such as measuring balanced or imbalanced triads by evaluating the mean contribution of one triad proportional to the difference between the number of balanced and imbalanced triads, with ranges between $-1$ and $+1$ corresponding to a balanced and an imbalanced structure, respectively \cite{marvel}. The dynamics of Heider’s social balance is a relatively recent research topic. Antal et al. proposed a discrete-time model in which links change sign with the aim of balancing imbalanced triads in a fully connected network. In this model, a network converges into a steady state with balanced triads\cite{Antal}. Kulakowski et al. proposed a continuous-time model under global adjustment to explain the process of transforming a fully connected network into a balanced one in finite time \cite{kulakowski2005heider}. Abell and Ludwig showed the process of balance in an incomplete signed network by variations in the number of positive links and the tolerance to imbalance \cite{abell2009structural}.
	 
	Previous works using statistical analysis inspired by the Ising model to signed networks go back to Refs. \cite{newman2000mean,albert2002statistical,park2005solution,dorogovtsev2008critical}.
	In those works, the spin variables are subject to interactions (links sign) and the mean-field approximation is used as an accurate solution, with regard to the symmetry of a fully connected network. Here, we build on the work of \cite{newman2000mean} and focus on the particular effect of temperature on the dynamics of balance theory. Several other studies have investigated balance theory following the same approach
	 \cite{belaza2017statistical,du2018reversing,belaza2019social,kirkley2019balance}, although there are some differences as well. Belaza et al. have written a Hamiltonian with three-body, two-body, and one-body interactions to study balance theory in political networks \cite{belaza2017statistical}.

	In this work, we use the balance theory Hamiltonian in a many-body system described by Boltzmann-Gibbs statistics. In terms of balance theory, imbalanced triads are sources of stress and therefore tend to be avoided by individuals when they update their personal relationships. Therefore, we introduce the temperature as a randomness that enables the network to keep some of those triads. First, we give a solution for our model based on a mean-field approximation that leads to a first-order phase transition among the variations of temperature. Then, we test the accuracy of the analytic results by using Monte Carlo simulation. We show how the converged state of a network can be changed by the temperature.

	\section{Model and Material}
	\label{sec-res}

	In this section, we consider the effects of temperature on the dynamics of social networks, focusing on the triadic relations, and we discuss how this may be linked to the system stability.
	
	The definition of structural balance relies on
	the concept of tensions in a network of individuals whose relations represent friendship and animosity (positive and negative). The balance theory includes a tendency toward reduced tensions, by changing the sign of links and thus increasing balanced triads, but we are interested to see how much tension  the system can tolerate. 
	
	\textcolor{black}{
	To describe the dynamics mechanism of the structural balance, we explain the update rules devoted to evolving networks. At each update event, we randomly select a link and switch its sign  to increase the total number of balanced triads. After an update step, some of the imbalanced triads become balanced. Note that the overall number of imbalanced triads cannot increase in an update event.
	A finite network falls in a balanced state where no imbalanced triads remain (except the jammed states).}
	
	\textcolor{black}{
	However, most of the previous works were devoted to analyzing the dynamics of networks by changing the link values or structure to understand tendencies towards or away from balance, without considering temperature. Antal \cite{Antal} proposed a formulation of balance theory in terms of energy with a focus on explaining why the systems do not necessarily evolve to a balanced state and are trapped in the so-called jammed states, i.e., the local minima. 
	We claim, via our study, that people generally show different levels of tolerance. Therefore, we explore the system dynamics under the new update rules based on the increase of imbalanced triads permitted by the system, which occur with some non-zero probability.} 

	\textcolor{black}{For this reason, we use the Boltzmann distribution which gives us the ability to calculate the probability and determine how much we can increase imbalanced triads at a certain temperature. As we can see, this will mean that by changing temperature we can allow more imbalanced triads to remain and so impose more tensions.} 
	
    \textcolor{black}{
	From a mathematical point of view, the newly introduced $"T"$ can be interpreted as a randomness of the social process. By social processes, we mean the ways in which individuals interact and modify their relationships to resolve tensions. 
	So the present structural balance is a special case of this model, in the zero-temperature non-equilibrium evolution through steady states. We further show that the change in temperature affects the Boltzmann distribution significantly, and the triads distribution depends on the temperature of the system.}

	\begin{figure}
		\hspace*{-1cm}
		\centering
		\includegraphics[width=8cm , height=2.35 cm ]{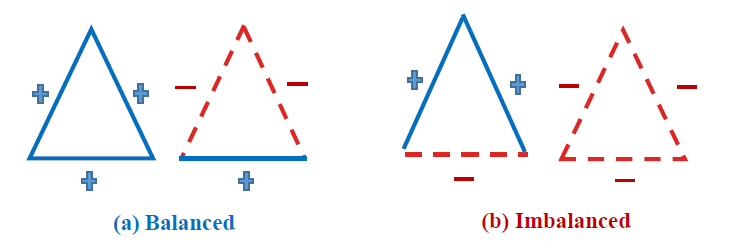}
		\caption{different types of triads according to structural balance}\label{fig1}
	\end{figure}

   In the structural balance, the units of analysis are two types of triads (balanced, imbalanced), presented in Fig.(\ref{fig1}). Assigning an energy $H(x)$ to each microstate of the system, we can model the probability distribution of a canonical system using the Boltzmann formula  $p(x) \sim e^{-\beta H(x)}$ where $"\beta=1/T"$ has the role of an inverse temperature. \textcolor{black}{$H(x)$ is calculated based on the number of balanced and imbalanced triads, and the more imbalanced triads there are, the higher h(x) would be.}
    At high temperatures, the system cannot move toward the balanced states, so that all type of triads presented in Fig.(\ref{fig1}) occur with the same probability and they will appear \textcolor{black}{randomly}. On the contrary, for low temperatures, the system will be in the lowest possible energy state with all balanced triads.

	In the next section, we present the Hamiltonian formulation with the description of a mean-field approach that shows that the model possesses a classic first-order phase transitions between two balanced and imbalanced states.

	\section{Hamiltonian and Mean-Field solution}

	In the present study, we focus on the case of fully connected networks, which usually are relatively easier to deal with in a theoretical treatment when mean-field methods are employed. Let's first specify the variables involved in the system. A social network is represented by a fully connected graph, which consists of a set of nodes and edges between them. The Hamiltonian of this network, according to the structural balance theory, is \cite{Antal}:
	
	 \begin{align}\label{eq:1}
	 H=-\sum_{i>j>k} S_{ij}S_{jk}S_{ki}
	 \end{align}
	
	Where $ S_{ij}=\{\pm 1 \}$ is the sign of the edge between nodes $i$ and $j$, and it encodes the relationship (friend or enemy) between them.
	We can analytically solve the Hamiltonian Eq.\eqref{eq:1} by using the mean-field solution.
	Let $H_{ij}$ be the sum of all terms in the Hamiltonian Eq.\eqref{eq:1}, that involve $S_{ij}$, and let $H^{'}$ be the remaining terms related to other edges, so that the Hamiltonian can be written as follows:
	
	\begin{equation}\label{eq:2}
	\begin{aligned}
	&H = H_{ij}+H^{'}\\&
	H_{ij}=-S_{ij} \sum_{k \neq i,j} S_{jk}S_{ki}
	\end{aligned}
	\end{equation}

To calculate the mean value of $S_{ij}$, we need to define the probability as (Appendix A):
\begin{equation}\label{eq:2-1}
\begin{aligned}
&\langle S_{ij} \rangle=\sum_{S_{ij}=\{\pm 1\}}P(S_{ij})S_{ij}\\
\end{aligned}
\end{equation}

We have the mean value of $S_{ij}$ as:
	\begin{equation}\label{eq:3}
	\begin{aligned}
	\langle S_{ij} \rangle&=P(S_{ij}=1)*(1)+P(S_{ij}=-1)*(-1) =  \\&
	\Big\langle \frac{e^{-\beta H_{ij} (S_{ij}=1)}-e^{-\beta H_{ij} (S_{ij}=-1)} }{e^{-\beta H_{ij} (S_{ij}=1)}+e^{-\beta H_{ij} (S_{ij}=-1)}}\Big\rangle = \\&
	\Big\langle \frac{e^{\beta \sum_{k \neq i,j} S_{jk}S_{ki}}-e^{-\beta  \sum_{k \neq i,j} S_{jk}S_{ki}}}{e^{\beta \sum_{k \neq i,j} S_{jk}S_{ki}}+e^{-\beta  \sum_{k \neq i,j} S_{jk}S_{ki}}}\Big \rangle=\\&
	\Big\langle \tanh (\beta  \sum_{k \neq i,j} S_{jk}S_{ki})\Big \rangle.
	\end{aligned}
	\end{equation}

    Here, $\langle ...\rangle$ indicates an ensemble average over terms that are involved in the $H^{'}$ part of the Hamiltonian.
   
	The two-body term in the Hamiltonian can be interpreted as a force term that attempts to “homogenize” the relations in the triad.
	First, we need to define the Hamiltonian $H(S_{ik},S_{kj})$ as:
	\begin{equation}\label{eq:4}
	\begin{aligned}
	H_{ik,kj} &=-S_{ik}( \sum_{l \neq i,j,k}S_{il}S_{lk}  )-S_{kj}(\sum_{l \neq i,j,k}S_{kl}S_{lj})	-S_{ik}S_{kj}S_{ji}\\&
	=-S_{ik}\langle S_{i}|S_{k} \rangle  -S_{kj}\langle S_{k}|S_{j} \rangle
	-S_{ik}S_{kj}S_{ji}
	\end{aligned}
	\end{equation}
	
	By following the above mentioned steps, we can write the mean value $\langle S_{ik}S_{kj} \rangle$:

    \begin{widetext}
	\begin{equation} \label{eq:5}
	\begin{aligned}
	&\langle S_{ik}S_{kj} \rangle=
	\\&P(S_{ik}, S_{kj}=1)*(1) +P(S_{ik}=1,S_{kj}=-1)*(-1)
	 +P(S_{ik}=-1, S_{kj}=1)*(-1)+P(S_{ik},S_{kj}=-1)*(1)=\\&
\Big \langle \frac{e^{-\beta H_{ik,kj}(S_{ik},S_{kj}=1)}-e^{-\beta H_{ik,kj}(S_{ik}=1,S_{kj}=-1)}
 -e^{-\beta H_{ik,kj}(S_{ik}=-1,S_{kj}=1)}+e^{-\beta H_{ik,kj}(S_{ik},S_{kj}=-1)}}{
    e^{-\beta H_{ik,kj} (S_{ik},S_{kj}=1)}
    	+e^{-\beta H_{ik,kj}(S_{ik}=1,S_{kj}=-1)}
    	+e^{-\beta H_{ik,kj} (S_{ik}=-1,S_{kj}=1)}
    	+e^{-\beta H_{ik,kj} (S_{ik},S_{kj}=-1)} }\Big\rangle
	\end{aligned}
	\end{equation}	
	\end{widetext}

Depending on the sign of $S_{ik}$ and $S_{kj}$, we have four different equations as follow:\\

\begin{equation}\label{eq:6}
\begin{aligned}
	&H_{ik,kj}(S_{ik},S_{kj}=1) =-\langle S_{i}|S_{k} \rangle -\langle S_{k}|S_{j} \rangle -S_{ij}\\&
	H_{ik,kj}(S_{ik}=1,S_{kj}=-1) =-\langle S_{i}|S_{k} \rangle +\langle S_{k}|S_{j} \rangle +S_{ij}\\&
	H_{ik,kj}(S_{ik}=-1,S_{kj}=1)=\langle S_{i}|S_{k} \rangle -\langle S_{k}|S_{j} \rangle +S_{ij}\\&
	H_{ik,kj}(S_{ik},S_{kj}=-1)=\langle S_{i}|S_{k} \rangle +\langle S_{k}|S_{j} \rangle -S_{ij}
\end{aligned}
\end{equation}
		
	\begin{figure*}
	\centering
	\subfloat[]{{\includegraphics[width=8cm, height=6cm]{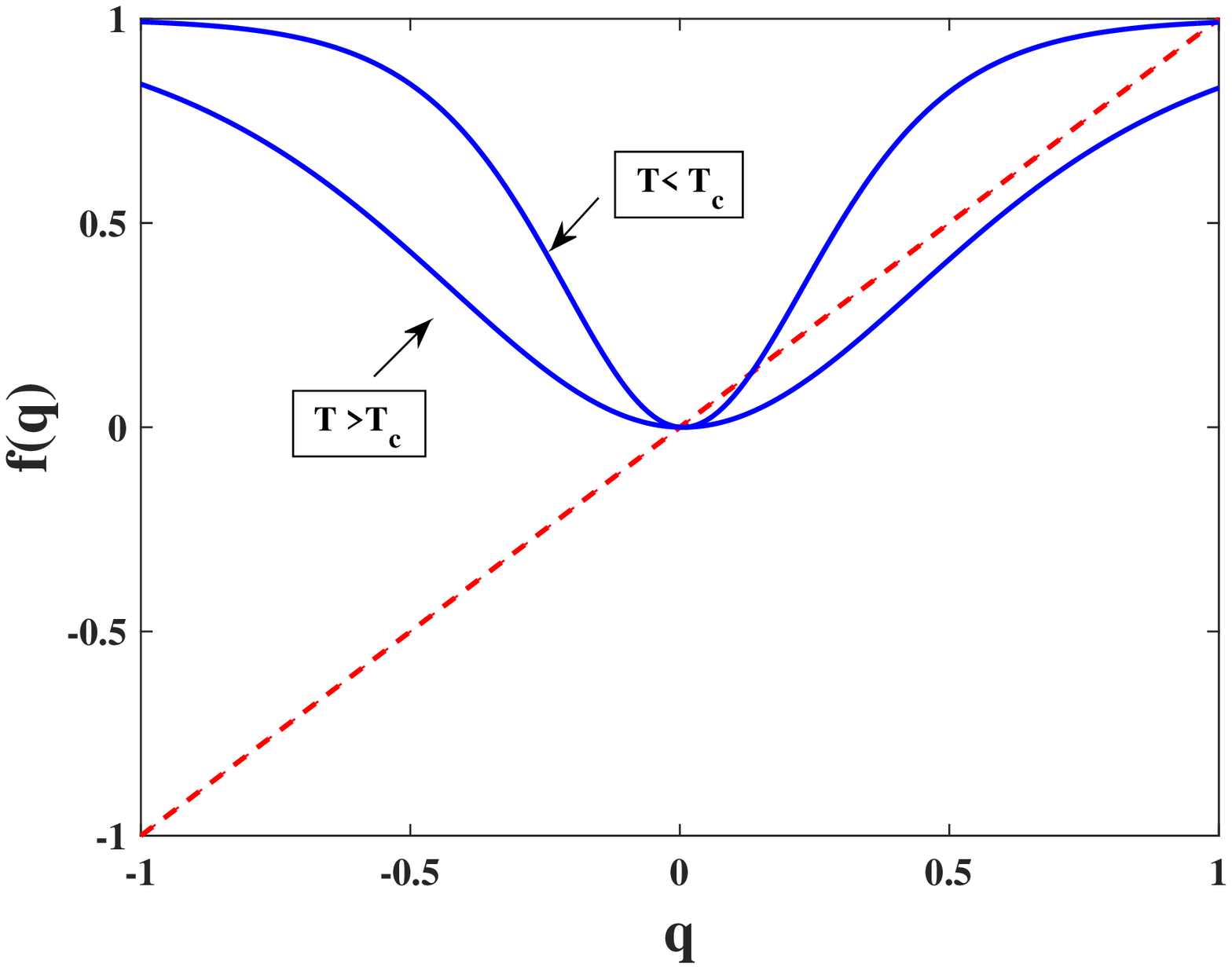}}}%
	\qquad
	\subfloat[]{{\includegraphics[width=8cm,height=6cm]{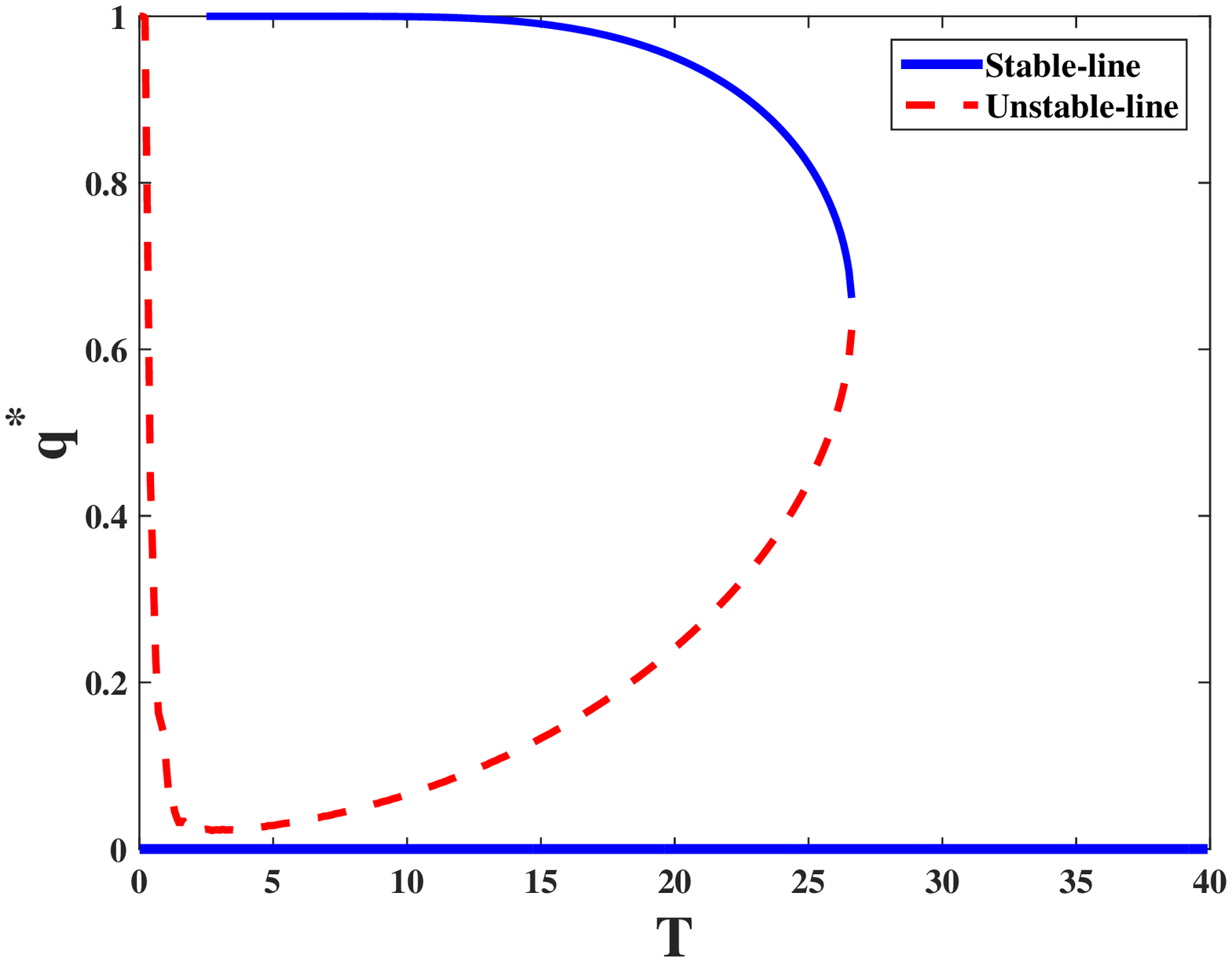} }}%
	\caption{(a) Graphical analysis of Eq.(\ref{eq:9}) for N=50 and in different temperatures $(T=15, T=28)$. When $T >T_{c}(T_{c}=26.2)$ there is a single stable fixed point. When $T <T_{c}$ there are three fixed points and the inner one is unstable. The value of $T$ determines the number of fixed points.
		(b) Bifurcation diagram showing the fixed points ($q^{*}$) as a function of temperature ($T$).}%
	\label{fig2}%
\end{figure*}	
At least,by substituting Eq.(\ref{eq:6}) in Eq.(\ref{eq:5}):
\noindent

\begin{widetext}	
	\begin{equation}\label{eq:8}
	 \begin{aligned}
	&\langle S_{ik}S_{kj} \rangle =\\&
	\scalebox{0.95}{$\Big \langle  \frac{e^{-\beta (N-3)(-\langle S_{i}|S_{k} \rangle -\langle S_{k}|S_{j} \rangle ) +\beta\langle S_{ij} \rangle}
	-e^{-\beta (N-3)(-\langle S_{i}|S_{k} \rangle +\langle S_{k}|S_{j} \rangle )- \beta \langle S_{ij} \rangle}-e^{-\beta (N-3)(\langle S_{i}|S_{k} \rangle -\langle S_{k}|S_{j} \rangle )- \beta \langle S_{ij} \rangle}
	+e^{-\beta (N-3)(\langle S_{i}|S_{k} \rangle +\langle S_{k}|S_{j} \rangle )+\beta \langle S_{ij} \rangle}
}{e^{-\beta (N-3)(-\langle S_{i}|S_{k} \rangle -\langle S_{k}|S_{j} \rangle )+\beta \langle S_{ij} \rangle}
		+e^{-\beta (N-3)(-\langle S_{i}|S_{k} \rangle +\langle S_{k}|S_{j} \rangle )-\beta \langle S_{ij} \rangle}+e^{-\beta (N-3)(\langle S_{i}|S_{k} \rangle -\langle S_{k}|S_{j} \rangle )-\beta \langle S_{ij} \rangle}+e^{-\beta (N-3)(\langle S_{i}|S_{k} \rangle +\langle S_{k}|S_{j} \rangle )+\beta \langle S_{ij} \rangle}}\Big \rangle$}
			\end{aligned}
		\end{equation}
\end{widetext}

   Now we set up an equation for $\langle S_{ik}S_{kj} \rangle$ by applying the mean-field approximation to Eq.(\ref{eq:8}), which involves replacing the variables with their ensemble averages, which in this case means $ S_{ik}S_{kj} \to q \equiv \langle S_{ik}S_{kj} \rangle$. Here, we have assumed that the total averages in Eqs.(\ref{eq:3}) and (\ref{eq:5}) can be approximated by averaging over internal terms \cite{park2005solution}(for more details see Appendix A). Defining also $p \equiv  \langle S_{ij} \rangle$, we have:

  \begin{equation}
  \begin{aligned}
   &p=\tanh (\beta {(N-2)}q)\\&
   q=\left[\frac{e^{-\beta (N-3)(-2q)}-2e^{-\beta (2p)}+e^{-\beta (N-3)(2q)}}
   {e^{-\beta (N-3)(-2q)}+2e^{-\beta (2p)}+e^{-\beta(N-3)(2q)}}\right]
   \end{aligned}
   \end{equation}

To solve these two equations with two unknowns, we use a self consistency condition on $q$ by substituting $p$ into $q$:

	\begin{equation}\label{eq:9}
	\begin{aligned}
	\scalebox{0.99}{$q=\left[\frac{
		e^{-\beta (N-3)(-2q)}-2e^{-2\beta \tanh (\beta (N-2)q)}+e^{-\beta (N-3)(2q)}}
	{e^{-\beta (N-3)(-2q)}+2e^{-2\beta\tanh (\beta  (N-2)q)}+e^{-\beta(N-3)(2q)}}\right]
	=f(q)$}
	\end{aligned}
	\end{equation}
	\\
	
	 In Fig.(\ref{fig2}), we show a plot of the forms $y=q$ and $y=f(q)$ as functions of $ q $, and we discuss the bifurcation that appears in the system when varying the temperature. 
	 The intersections of the line and curve give the solutions of Eq.(\ref{eq:9}). If the system is in $T >T_{c}$, there is only one stable fixed point $q^{*}=0$ that corresponds to a random state of the system. In $T=T_{c}$, two additional fixed points appear, one stable and one unstable. In $T<T_{c}$, the new stable $q^{*}$ will grow rapidly until it collides with the $q^{*}=1$, which  corresponds to a balanced state of the system.
    
	The bifurcation diagram for the dynamical system with $N=50$ in Eq.\eqref{eq:9}, shows a "blue sky" bifurcation, and the system undergoes a phase transition between the possible solutions. Suppose we cool down the system slowly, so that $q^{*}$ flows until the critical point $T_{c}$ is approached and then a saddle-node bifurcation with a pair consisting of an unstable and a stable point takes place. Hence, besides the existing fixed point $q^{*}=0$, two other fixed points will be created: one stable and one unstable. The unstable fixed point separates the basins of attraction of the two stable fixed points.
	
	Finally, we calculate the mean-field equation for $\langle S_{ik} S_{kj} S_{ji}\rangle$ which gives the mean contribution of one triad or the mean triad energy and a sense of stability in the network (see Appendix B):

\begin{widetext}
	\begin{equation}\label{eq:10}
	\begin{aligned}
	\langle S_{ik}S_{kj}S_{ji} \rangle&= \left[\frac{
			e^{\beta(N-3)(3q)+\beta}-3e^{\beta(N-3)(q)- \beta}{+3e^{\beta(N-3)(-q)+\beta}-e^{\beta(N-3)(-3q)-\beta}}}{{
				e^{\beta(N-3)(3q)+\beta }+3e^{\beta(N-3)(q)-\beta}{+3e^{\beta(N-3)(-q)+\beta}+e^{\beta(N-3)(-3q)-\beta}}}}\right]
	\end{aligned}
	\end{equation}
\end{widetext}

\section{SIMULATIONS}
	
	To describe how networks are affected by the temperature, and to obtain results demonstrating our main point, we start out with a fully connected network with N nodes and a random configuration. The procedure we follow is a Metropolis algorithm based on the structural balance Hamiltonian \cite{Antal} to generate paths to reach the stable state.
In each step for a given temperature, we select a link randomly. Then, using Eq.\eqref{eq:1} the energy difference $ \Delta E = E_{n+1}-E_{n}$ between the new configurations $E_{n+1}$ and that belonging to the prior $E_{n}$ is calculated. The chosen link is flipped if $ \Delta E <0$ and also if $ \Delta E>0 $ with a probability equal to $"\exp{\left(-\beta \Delta E\right)}"$ where $"\beta"$ is equal to the inverse of the temperature.
	
\begin{figure}[H]
 	\centering
 	\includegraphics[width=8cm,height=6cm,angle=0]{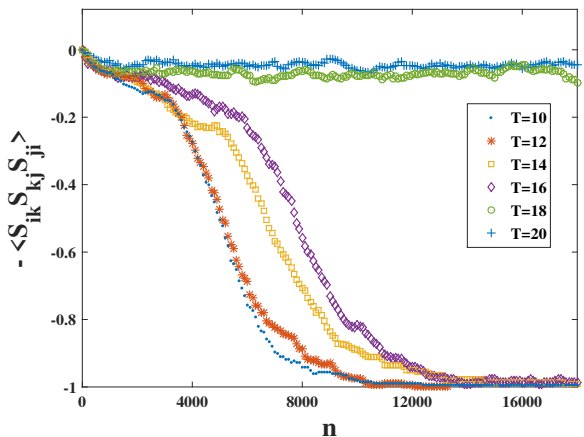}
 	\caption{The mean triad energy ($-\langle S_{ik} S_{kj} S_{ji}\rangle$) versus Monte Carlo steps (n) for different value of temperatures in a fully connected network with N=50.}\label{fig3}
 \end{figure}

   Repeating this procedure for different values of temperature reveals the diagram shown in Fig.\ref{fig3}. We see that after a specific temperature the system does not reach  the balanced state and stays in its imbalanced state with energy equal to zero. This implies that temperature has important effects on convergence. 
 
 	\begin{figure*}
 	\centering
 	\subfloat[]{{\includegraphics[width=8cm,height=6.2cm]{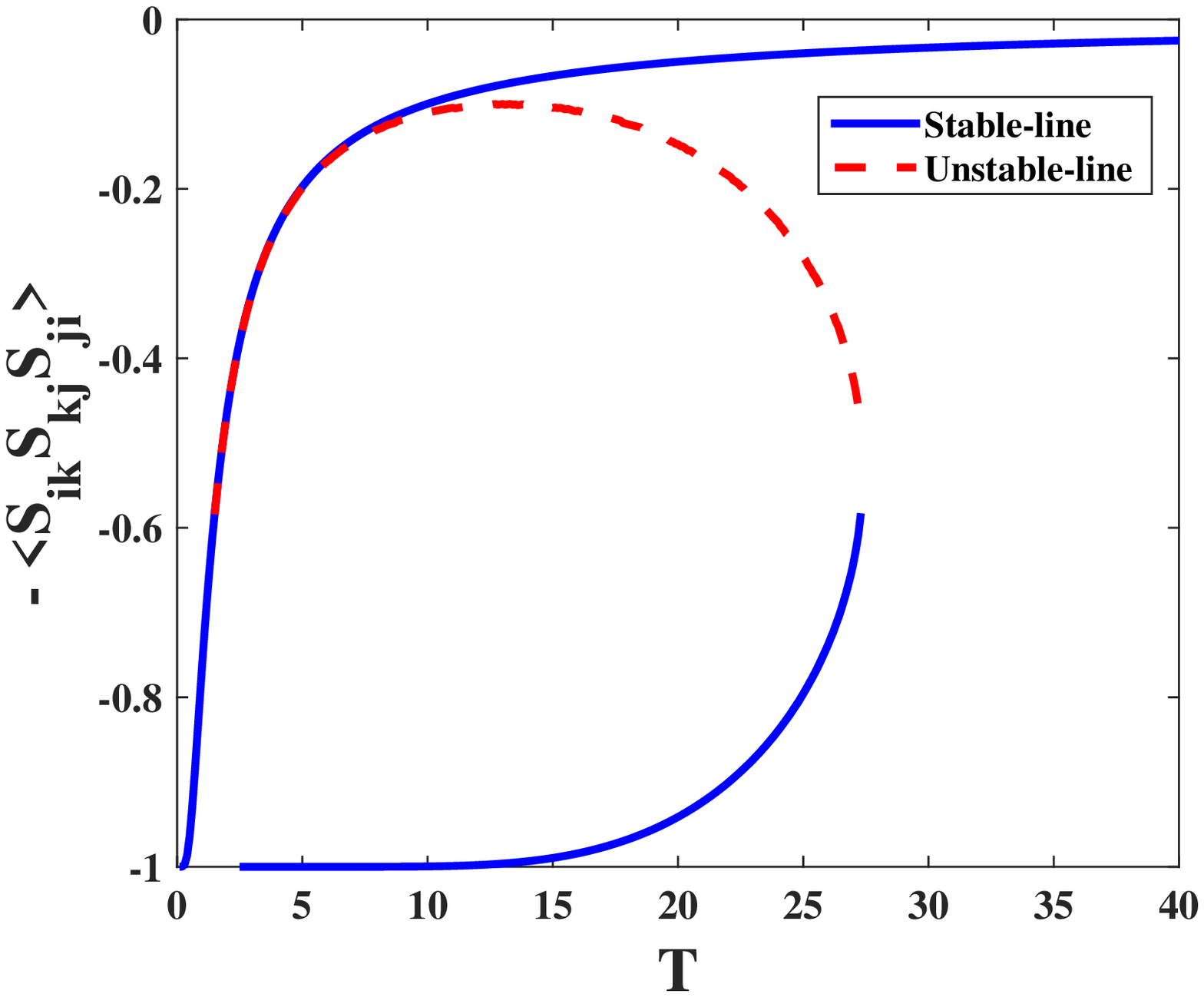}}}%
 	\qquad
 	\subfloat[]{{\includegraphics[width=7.8cm,height=5.9 cm]{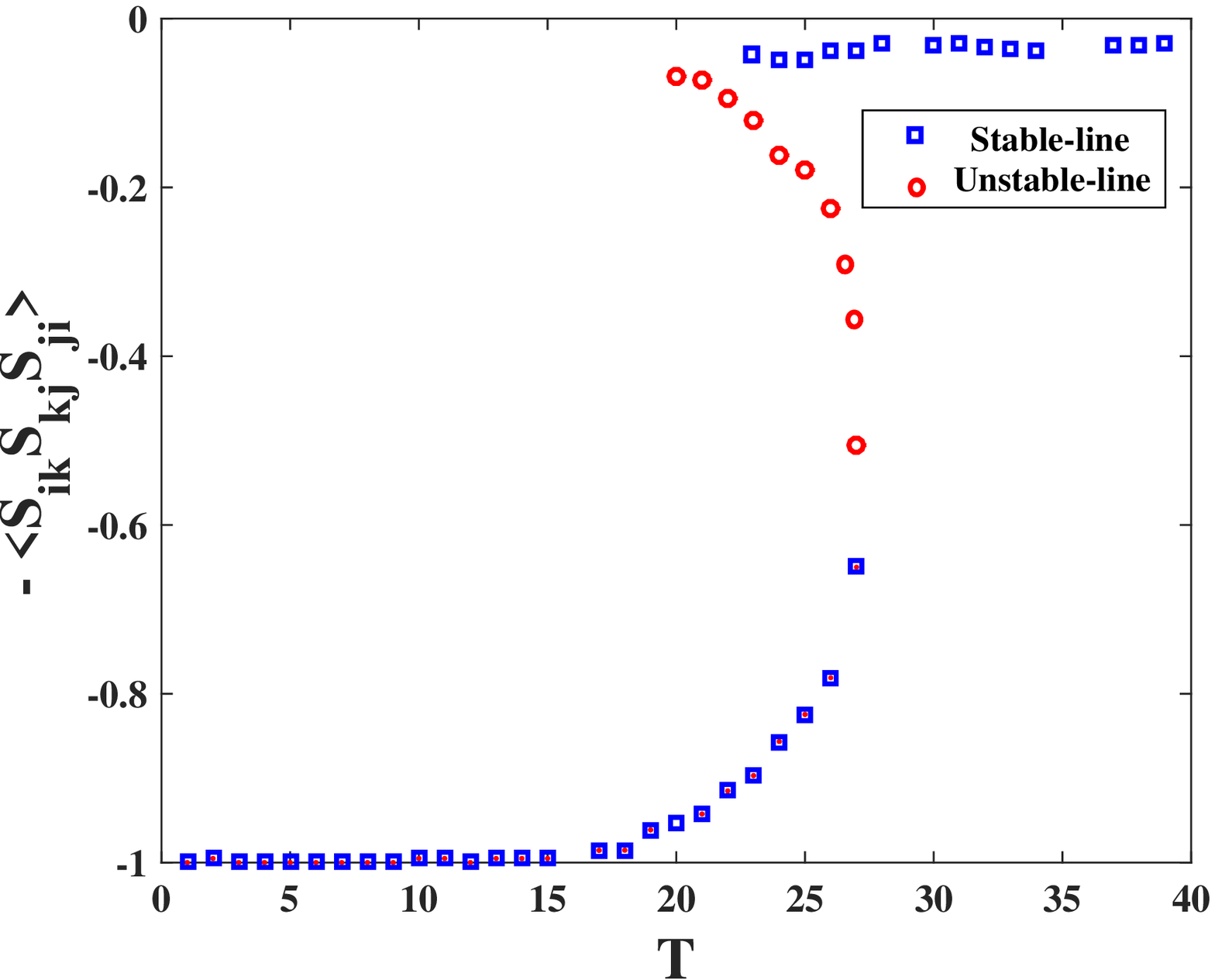} }}%
 	\caption{(a) Bifurcation diagram for the first-order phase transition in Eq.(\ref{eq:10}) as a function of temperature. The solid blue curve represents stable fixed points, while the dashed red curve is the unstable one. (b) 
 	Monte Carlo simulation from several initial states with different mean triad energies. To generate these initial states, we build a balanced network with all positive links and then randomly change the sign of some links to negative.
 		There is a first-order phase transition at $T=T_{c}$, where the mean triad energy jumps from $-1$ to $0$. This is known as a "blue-sky" bifurcation. The results correspond to a fully connected network with N=50.}%
 	\label{fig4}%
 \end{figure*}
  In Fig.\ref{fig4}-(a), we can see the mean triad energy based on our analytical results, Eq.\eqref{eq:10}, which shows the first-order phase transition and thermal hysteresis. We see that  bistability exists, and as a definite temperature $(T=T_{c})$ the mean triad energy changes abruptly. This leads to a hysteresis phenomenon that is typical for a first-order phase transition.
  
  For comparison, the Monte Carlo simulation's result in Fig.\ref{fig4}-(b) represent the behavior of the same object based on our mean-field solution. 
  For each temperature, we start our Monte Carlo simulation from several initial states with different mean triad energies. To generate these initial states, we build a balanced network with all positive links and then randomly change the sign of some links to negative. After that, we evolve the networks to reach the steady state and take the average on all of them to obtain the mean triad energy.

   Throughout our simulations, some of them reach to the line with $-\langle S_{ik}S_{kj}S_{ji} \rangle=0$ and others to $-\langle S_{ik}S_{kj}S_{ji} \rangle=-1$. To make the unstable curve, we consider a state with specific mean triad energy in which all states with lower mean triad energies lead to $-\langle S_{ik}S_{kj}S_{ji} \rangle=-1$, and states with upper mean triad energies lead to $-\langle S_{ik}S_{kj}S_{ji} \rangle=0$.

  However, a discrepancy exists between theory and simulation for the prediction of phase transition points. One of the main reasons may be that near phase transition points the lifetime of one of the metastable states becomes short so that the metastable state cannot be fully sampled in the simulation.

  As we have shown in Fig.\ref{fig4}-(a), for $(T \sim 0)$, the only balanced configuration with $-\langle S_{ik}S_{kj}S_{ji} \rangle=-1$ is stable. For $T>T_{c}$, the only random configuration with $-\langle S_{ik}S_{kj}S_{ji} \rangle=0$ is stable. In other words, our system will have two fixed points, one with all balanced triads $-\langle S_{ik}S_{kj}S_{ji} \rangle=-1$ and another one with almost the same number of imbalanced and balanced triads $(-\langle S_{ik}S_{kj}S_{ji} \rangle \sim 0)$. In the region $(0 \ll T<T_{c})$, the two metastable phases with $-\langle S_{ik}S_{kj}S_{ji} \rangle \sim -1 $ and $-\langle S_{ik}S_{kj}S_{ji} \rangle \sim 0 $ coexist, separated by an unstable state (dashed line). This solution shows the metastability of the system which is a state with both types of triads, and just changing one link in a triad can move it toward two other stable points.

  \begin{figure}
  	\centering
  	\includegraphics[width=8cm,height=6cm,angle=0]{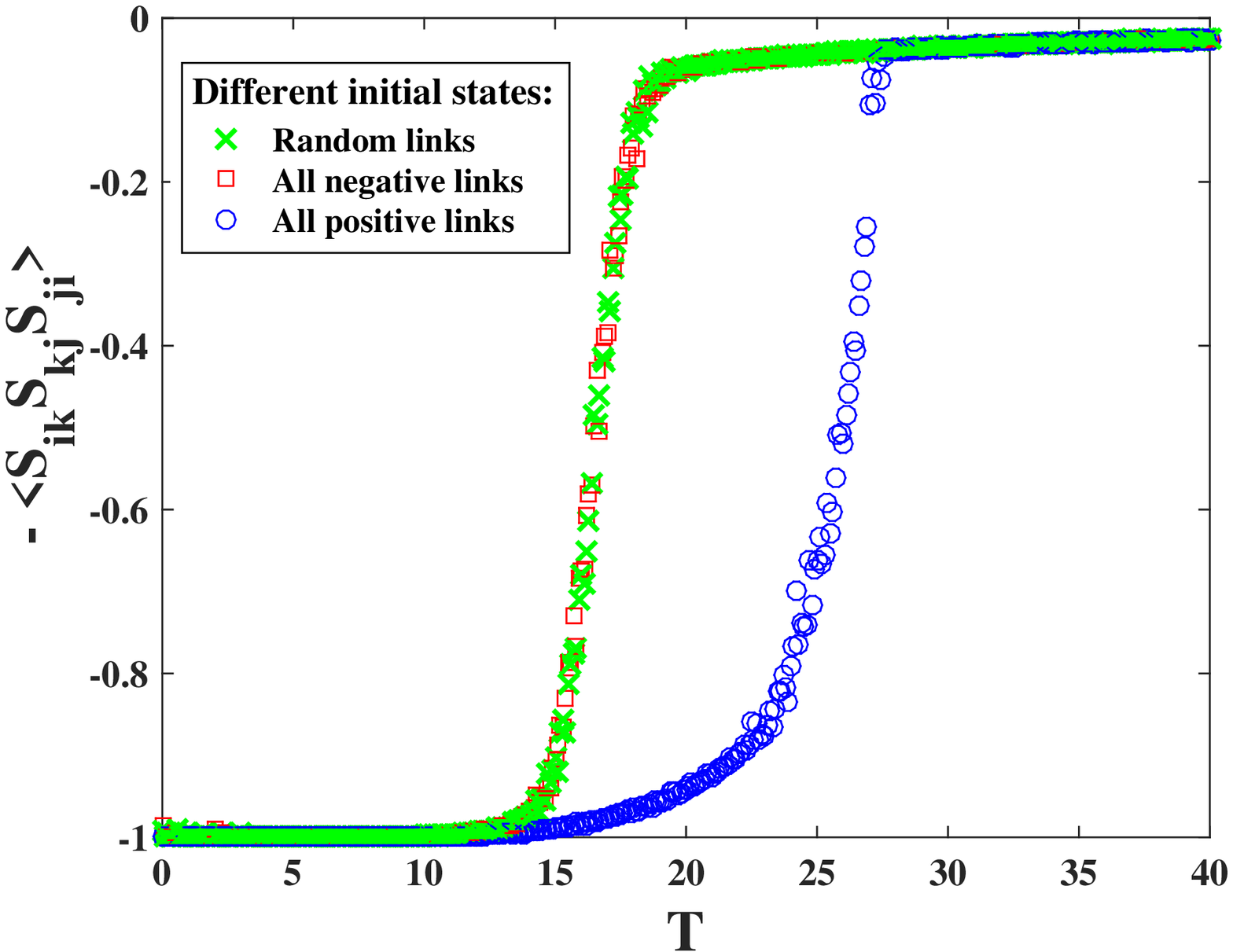}
  	\caption{The mean triad energy ($-\langle S_{ik}S_{kj}S_{ji} \rangle$) versus temperature (T) for different initial states differing in the percent of balanced triads. We consider a balanced initial state with all positive links ($-\langle S_{ik}S_{kj}S_{ji} \rangle=-1$)} and imbalanced initial state with all negative links ($-\langle S_{ik}S_{kj}S_{ji} \rangle=1$). In random initial state, we have almost equal positive and negative links ($-\langle S_{ik}S_{kj}S_{ji} \rangle \sim 0$).\label{fig5}
  \end{figure}

  To investigate the tendency to the balanced state corresponding to the coexistence region, we start our simulation with different configurations differing in the percentage of balanced triads (Fig.\ref{fig5}). We consider a balanced initial state with all positive links ($-\langle S_{ik} S_{kj}S_{ji}\rangle =-1$) and an imbalanced one with all negative links ($-\langle S_{ik}S_{kj}S_{ji}\rangle=1$). In random initial state, we have almost equal positive and negative links $-\langle S_{ik}S_{kj}S_{ji}\rangle\sim 0$. Fig.\ref{fig5} shows that trajectories starting from a balanced initial state lie into the balanced fixed point at $-\langle S_{ik} S_{kj}S_{ji}\rangle =-1$, while trajectories starting from random and imbalanced initial states go to the random fixed point at $-\langle S_{ik} S_{kj}S_{ji}\rangle =0$.

   By considering the temperature as a randomness of the
  	social process, regardless of the balanced state of triads, we could extend our results to society at large. Our findings suggest that there is a threshold, a critical temperature, for tolerating tensions, beyond which society splits into two separate groups with antagonistic relations, such as  political parties.
  	Hence, the first consequence of the phase transition is the non-gradual and abrupt phase change of society. In other words, society does not respond linearly to the tolerance level of its agents, i.e. temperature. It has a random phase and a polarized phase, which are separated by a critical temperature.
  \\
  Another consequence of our results comes from the first-order attribute of phase transition. If the system is quenched to temperatures below the critical point, which means the cooling down process is much faster than the time needed to reach thermal equilibrium, system stays in random phase. Therefore, from a sociological perspective, the tolerance level is below the critical point, but society is still in its random phase. This is a vulnerable state, because potentially, society will jump to a polarized phase.
  \\
  In addition, it shows a hysteresis phenomenon where increasing temperature does not return it to random phase, until the temperature goes beyond the critical point. Thus, controlling society's state and taking that out of its polarized phase would be a difficult job.\\
  
	\section{CONCLUSION}
	\label{sec-disc}

    To find the dynamical behavior of tensions in signed networks based on structural balance theory, we propose a model that takes into account the impacts of temperature. Based on this model, we analyze the dynamical process and provide a mean-field solution for identifying the threshold of tolerance in networks with signed relationships. The mean-field approximation reveals an abrupt (first-order) phase transition in the system's balanced states, as a function of temperature. It has a crucial consequence: after cooling a system to $T<T_c$, it may stay in its random state, but a little perturbation might take it to become unstable and resulting in a polarized (balanced) state.

    By performing Monte Carlo simulations for the different signed networks, we find that the theoretical and numerical results are in good agreement, confirming the correctness of our mean-field approximation. Furthermore, the networks' transition from an imbalanced state with tension to a steady state is not smooth, i.e., it involves an abrupt change. Increasing temperature does not mean a corresponding increase of tensions. In other words, even with increasing temperature, the system reaches a balanced state with a bipartite configuration.
	However, there is a critical temperature $T_{c}$ (the threshold of tolerance) beyond which there is no chance for structural balance.


\bibliographystyle{apsrev4-1}	
\bibliography{test}

\clearpage
  \appendix
  \begin{appendices}
  \color{black}
  	
  	\section{Partition Function Calculations}
  	
  	In this appendix, we have given the partition function calculations for the Hamiltonian Eq. (\ref{eq:1}) and introduce $h_{ij}$ as an external field on $ S_{ij}$,
  	
    \begin{equation}\label{eq:A0-1-1}
   \begin{aligned}
   &H_{ij}=-S_{ij}\sum_{S_{ij}=\pm 1} S_{jk}S_{ki}-S_{ij}h_{ij}.
   \end{aligned}
   \end{equation}
   
   Therefore, the partition function can be written as:
   
  	\begin{equation}\label{eq:A0-1}
  	\begin{aligned}
     Z&=\sum_{\{S\}} e^{- \beta H}=\sum_{\{S\}}e^{-\beta (H_{ij}+H^{'})}=\\&
     Z^{'}\sum_{S \neq S_{ij}}\frac{e^{- \beta H^{'}}}{Z^{'}} \sum_{S_{ij}=\{\pm 1\}} e^{- \beta H_{ij}}=\\&
     Z^{'} \langle \sum_{S_{ij}=\{\pm 1\}} e^{- \beta H_{ij}} \rangle_{Z^{'}}=\\&
     Z^{'} \langle \cosh (\beta \sum_{k \neq i,j} S_{jk}S_{ki}+\beta h_{ij}) \rangle_{Z^{'}}\\&
     \end{aligned}
     \end{equation}
     
     	Where $Z^{'} =\sum_{\{S\}}e^{-\beta H^{'}}$ is the partition function for the Hamiltonian $H^{'}$, and $\langle...\rangle_{Z^{'}}$ indicates an ensemble average over it. Expanding the $cosh$ in a power series, we get:
     \begin{equation}\label{eq:A0-1-01}
     \begin{aligned}
     Z&=Z^{'} \langle \sum^{\infty}_{n=0} \frac{\Big(\beta (\sum_{k \neq i,j} S_{jk}S_{ki}+h_{ij})\Big)^{2n}}{(2n)!} \rangle_{Z^{'}}=\\&
     Z^{'} \sum^{\infty}_{n=0} \frac{(\beta)^{2n} \langle (\sum_{k \neq i,j} S_{jk}S_{ki}+h_{ij}) ^{2n} \rangle_{Z^{'}}}{(2n)!}=
     \\&
     \scalebox{0.99}{$Z^{'} \sum^{\infty}_{n=0} \frac{(\beta)^{2n} \langle \sum_{m=0} C^{2n}_{m} ( \sum_{k \neq i,j} S_{jk}S_{ki})^{m} (\beta h_{ij}) ^{2n-m} \rangle_{Z^{'}}}{(2n)!}=$}
     \\&
    \scalebox{0.99}{$ Z^{'} \sum^{\infty}_{n=0} \frac{(\beta)^{2n} \sum_{m=0} C^{2n}_{m} \langle ( \sum_{k \neq i,j} S_{jk}S_{ki})^{m} \rangle_{Z^{'}} (\beta h_{ij}) ^{2n-m}}{(2n)!}=$}\\&
     \scalebox{1.04}{$Z^{'} \sum^{\infty}_{n=0} \frac{(\beta)^{2n} \sum_{m=0} C^{2n}_{m} \langle  \sum_{k \neq i,j} S_{jk}S_{ki}\rangle_{Z^{'}}^{m}  (\beta h_{ij}) ^{2n-m}}{(2n)!}=$}\\&
     Z^{'} \sum^{\infty}_{n=0} \frac{(\beta \langle \sum_{k \neq i,j} S_{jk}S_{ki}\rangle_{Z^{'}}+\beta h_{ij}) ^{2n} }{(2n)!}=\\&
     Z^{'} \cosh (\beta \langle\sum_{k \neq i,j} S_{jk}S_{ki}\rangle_{Z^{'}}+\beta h_{ij})
      \end{aligned}
     \end{equation}
     
     
     Where in the last line, based on a mean-field approximation, we have made the assumption that the correlation of link status more than two can be approximated as a power of $\langle S_{jk}S_{ki}\rangle_{Z^{'}}=q$. 
     
     Now, we can write the free energy as:
     


  	\begin{equation}\label{eq:A0-2}
  	\begin{aligned}
  	F&=-\beta^{-1} \ln Z= \\
  	&=-\beta^{-1} \ln (Z^{'} \cosh (\beta \langle\sum_{k \neq i,j} S_{jk}S_{ki}\rangle_{Z^{'}}+\beta h_{ij}))=\\&
  	-\beta^{-1} \ln (Z^{'})-\beta^{-1} \ln ( \cosh (\beta \langle\sum_{k \neq i,j} S_{jk}S_{ki}\rangle_{Z^{'}}+\beta h_{ij}))
  	\end{aligned}
  	\end{equation}
    
  The derivative of free energy with respect to external field, expresses the mean value of $S_{ij}$:
  
  	\begin{equation}\label{eq:A0-3}
  	\begin{aligned}
  	\langle S_{ij} \rangle&=-\frac{\partial F}{\partial h_{ij}}\bigm|_{h_{ij}=0}
  	=-\frac{\partial (-\beta^{-1} \ln Z )}{\partial h_{ij}}\bigm|_{h_{ij}=0}\\
  	&=\frac{ \sinh (\beta \langle\sum_{k \neq i,j} S_{jk}S_{ki}\rangle_{Z^{'}}+\beta h_{ij})}{ \cosh (\beta \langle\sum_{k \neq i,j} S_{jk}S_{ki}\rangle_{Z^{'}}+\beta h_{ij})}\Bigm|_{h_{ij}=0}\\
  	&= \tanh (\beta \langle\sum_{k \neq i,j} S_{jk}S_{ki}\rangle_{Z^{'}})
  	\end{aligned}
  	\end{equation}
  
  	For the two-body term, we repeat all these steps:
  
  	  	\begin{equation}\label{eq:A0-4}
  	\begin{aligned}
  	Z&=\sum_{{\{S\}}} e^{- \beta H}=\sum_{\{S\}}e^{-\beta (H_{ij}+H^{'})}=\\&
  	Z^{'}\sum_{S \neq S_{ik},S_{kj}}\frac{e^{- \beta H^{'}}}{Z^{'}} \sum_{S_{ik},S_{kj}=\pm 1} e^{- \beta H_{ik,kj}}=\\&
  	Z^{'} \langle \sum_{S_{ik},S_{kj}=\pm 1} e^{- \beta H_{ik,kj}} \rangle_{Z^{'}}
  	\end{aligned}
  	\end{equation}
  	\\
  	Here,
  	\ the Hamiltonian $H_{ik,kj}$is defined as:
  		\begin{equation}\label{eq:A0-4-0}
  	\begin{aligned}
  	H_{ik,kj}&=-S_{ik} (\sum_{l \ne i,j,k} S_{il}S_{lk}+h_{ik})-S_{kj} (\sum_{l \ne i,j,k} S_{kl}S_{lj}+h_{kj})\\
  	&-S_{ik}S_{kj}(S_{ij}+h_{ik,kj})
  	\end{aligned}
  	\end{equation}
  	
  	Here, $h_{ik,kj}$ is the two-body term external field. Depending on the sign of $S_{ik}$ and $S_{kj}$, we have four different equations as follow:
  	
  	\begin{equation}\label{eq:A0-5}
  	\begin{aligned}
  	&H_{ik,kj}(S_{ik},S_{kj}=1) =\\&
  	-\sum_{l \ne i,j,k} S_{il}S_{lk}-h_{ik}-\sum_{l \ne i,j,k}S_{kl}S_{lj} -h_{kj} -S_{ij}-h_{ik,kj}\\&
    H_{ik,kj}(S_{ik}=1,S_{kj}=-1) =\\&
    -\sum_{l \ne i,j,k}S_{il}S_{lk} -h_{ik}+\sum_{l \ne i,j,k} S_{kl}S_{lj} +h_{kj} +S_{ij}+h_{ik,kj}\\&
  	H_{ik,kj}(S_{ik}=-1,S_{kj}=1)=\\&
  	\sum_{l \ne i,j,k}S_{il}S_{lk} +h_{ik}-\sum_{l \ne i,j,k} S_{kl}S_{lj}-h_{kj} +S_{ij}+h_{ik,kj}\\&
  	H_{ik,kj}(S_{ik},S_{kj}=-1)=\\&
  	\sum_{l \ne i,j,k} S_{il}S_{lk} \sum_{l \ne i,j,k}+h_{ik}+\sum_{l \ne i,j,k} S_{kl}S_{lj}+h_{kj} -S_{ij}-h_{ik,kj}
  	\end{aligned}
  	\end{equation}\\
  		
  		If we now substitute in here the expressions for $H_{ik,kj}$, we get:

  		\begin{widetext}
  	\begin{equation}\label{eq:A0-6-2}
  	\begin{aligned}
  	Z&=
  	Z^{'} \langle  e^{-\beta (	-\sum_{l \ne i,j,k} S_{il}S_{lk}-h_{ik}-\sum_{l \ne i,j,k}S_{kl}S_{lj} -h_{kj} -S_{ij}-h_{ik,kj}}
  	+e^{-\beta(-\sum_{l \ne i,j,k}S_{il}S_{lk} -h_{ik}+\sum_{l \ne i,j,k} S_{kl}S_{lj} +h_{kj} +S_{ij}+h_{ik,kj})}\\&
  	+e^{-\beta(\sum_{l \ne i,j,k}S_{il}S_{lk} +h_{ik}-\sum_{l \ne i,j,k} S_{kl}S_{lj}-h_{kj} +S_{ij}+h_{ik,kj})}
  	+e^{-\beta (\sum_{l \ne i,j,k} S_{il}S_{lk} \sum_{l \ne i,j,k}+h_{ik}+\sum_{l \ne i,j,k} S_{kl}S_{lj}+h_{kj} -S_{ij}-h_{ik,kj})} \rangle_{Z^{'}} 	
  	\end{aligned}
  	\end{equation}
  		\end{widetext}
  	
  	Considering the same approximation as in previous calculations:
  	
  		\begin{equation}\label{eq:A0-6}
  		\begin{aligned}
  		Z&=Z^{'}
  		(e^{-\beta (N-3)(-\langle S_{i}|S_{k} \rangle_{Z^{'}} -h_{ik}-\langle S_{k}|S_{j} \rangle_{Z^{'}} -h_{kj} -S_{ij}-h_{ik,kj})}\\&
  	+e^{-\beta (N-3)(-\langle S_{i}|S_{k} \rangle_{Z^{'}} -h_{ik}+\langle S_{k}|S_{j} \rangle_{Z^{'}} +h_{kj} +S_{ij}+h_{ik,kj})}\\&
  			+e^{-\beta (N-3)(	\langle S_{i}|S_{k} \rangle_{Z^{'}} +h_{ik}-\langle S_{k}|S_{j} \rangle_{Z^{'}} -h_{kj} +S_{ij}+h_{ik,kj})}\\&
  			+e^{-\beta (N-3)(\langle S_{i}|S_{k} \rangle_{Z^{'}} +h_{ik}+\langle S_{k}|S_{j} \rangle_{Z^{'}} +h_{kj} -S_{ij}-h_{ik,kj})})
  		\end{aligned}
  		\end{equation}

  	So, the free energy is proportional to:
  	
  	\begin{equation}\label{eq:A0-7}
  	\begin{aligned}
  	F&=-\beta^{-1} \ln Z=-\beta^{-1} \ln Z^{'}\\&
  	-\beta^{-1} \ln \Big(e^{-\beta (N-3)(-\langle S_{i}|S_{k} \rangle_{Z^{'}} -h_{ik}-\langle S_{k}|S_{j} \rangle_{Z^{'}} -h_{kj} -S_{ij}-h_{ik,kj})}\\&
  	+e^{-\beta (N-3)(-\langle S_{i}|S_{k} \rangle_{Z^{'}} -h_{ik}+\langle S_{k}|S_{j} \rangle_{Z^{'}} +h_{kj} +S_{ij}+h_{ik,kj})}\\&
  	+e^{-\beta (N-3)(	\langle S_{i}|S_{k} \rangle_{Z^{'}} +h_{ik}-\langle S_{k}|S_{j} \rangle_{Z^{'}} -h_{kj} +S_{ij}+h_{ik,kj})}\\&
  	+e^{-\beta (N-3)(\langle S_{i}|S_{k} \rangle_{Z^{'}} +h_{ik}+\langle S_{k}|S_{j} \rangle_{Z^{'}} +h_{kj} -S_{ij}-h_{ik,kj})} \Big)
  	\end{aligned}
  	\end{equation}

  	Taking now the derivative of the free energy Eq.(\ref{eq:A0-7}), we get the result:
  	
  		\begin{widetext}
  	\begin{equation}\label{eq:A0-8}
  	\begin{aligned}
  	&\langle S_{ik}S_{kj} \rangle=-\frac{\partial F}{\partial h_{ik,kj}}\bigm|_{h_{ik,kj}=0}
  	=-\frac{\partial (-\beta^{-1} \ln Z )}{\partial h_{ik,kj}}\bigm|_{h_{ik,kj}=0}=\\&
  	\scalebox{0.98}{$\Big \langle  \frac{e^{-\beta (N-3)(-\langle S_{i}|S_{k} \rangle -\langle S_{k}|S_{j} \rangle ) +\beta\langle S_{ij} \rangle}
  		-e^{-\beta (N-3)(-\langle S_{i}|S_{k} \rangle +\langle S_{k}|S_{j} \rangle )- \beta \langle S_{ij} \rangle}-e^{-\beta (N-3)(\langle S_{i}|S_{k} \rangle -\langle S_{k}|S_{j} \rangle )- \beta \langle S_{ij} \rangle}
  		+e^{-\beta (N-3)(\langle S_{i}|S_{k} \rangle +\langle S_{k}|S_{j} \rangle )+\beta \langle S_{ij} \rangle}
  	}{e^{-\beta (N-3)(-\langle S_{i}|S_{k} \rangle -\langle S_{k}|S_{j} \rangle )+\beta \langle S_{ij} \rangle}
  		+e^{-\beta (N-3)(-\langle S_{i}|S_{k} \rangle +\langle S_{k}|S_{j} \rangle )-\beta \langle S_{ij} \rangle}+e^{-\beta (N-3)(\langle S_{i}|S_{k} \rangle -\langle S_{k}|S_{j} \rangle )-\beta \langle S_{ij} \rangle}+e^{-\beta (N-3)(\langle S_{i}|S_{k} \rangle +\langle S_{k}|S_{j} \rangle )+\beta \langle S_{ij} \rangle}}\Big  \rangle_{Z^{'}}$}
  	\end{aligned}
  	\end{equation}
  		\end{widetext}
  	\color{black}
  	\section{Mean-Field solution for Three Interactions}
  	
  	We need to write all steps again for the three interactions. Let's start by Hamiltonian:
  	
  	\begin{equation}\label{eq:A1}
  	\begin{aligned}
  	H &=H_{ik,kj,ji}+H^{'}
  	\\H_{ik,kj,ji}&=H_{ik\neq j }+ H_{kj\neq i }+ H_{ji\neq k }
  	\end{aligned}
  	\end{equation}
  	
  	\begin{widetext}
  	\begin{equation}\label{eq:A2}
  	\begin{aligned}
  	H_{ik,kj,ji} &=-S_{ik}( \sum_{l \neq i,j,k}S_{il}S_{lk})-S_{kj}( \sum_{l \neq i,j,k}S_{kl}S_{lj})-S_{ji}( \sum_{l \neq i,j,k}S_{jl}S_{li})-S_{ik}S_{kj}S_{ji}
  	\end{aligned}
  	\end{equation}
  	\end{widetext}

\begin{widetext}
  	\begin{equation}\label{eq:A3}
  	\begin{aligned}
  	&\langle S_{ik}S_{kj}S_{ji} \rangle  =\\&
  	P(S_{ik}=1,S_{kj}=1,S_{ji}=1)*(1) 
  	+P(S_{ik}=-1,S_{kj}=1,S_{ji}=1)*(-1) 
  	+P(S_{ik}=1, S_{kj}=-1,S_{ji}=1)*(-1) \\&
  	+P(S_{ik}=1, S_{kj}=1,S_{ji}=-1)*(-1)
  	 + P(S_{ik}=-1,S_{kj}=-1,S_{ji}=1)*(1)
  + P(S_{ik}=-1,S_{kj}=1,S_{ji}=-1)*(1)	\\&
  	+ P(S_{ik}=1,S_{kj}=-1,S_{ji}=-1)*(1)+
  	P(S_{ik}=-1,S_{kj}=-1,S_{ji}=-1)*(-1),
  	\end{aligned}
  	\end{equation}
  \end{widetext}

  	So, we have:	
  	
  	\begin{equation}\label{eq:A4}
  	\begin{aligned}
  	\langle S_{ik}S_{kj}S_{ji} \rangle =\langle\frac{e^{-\beta H_{ik,kj,ji}(S_{ik},S_{kj},S_{ji})} }{\sum_{S_{ik},S_{kj},S_{ji} = \pm 1}e^{-\beta H_{ik,kj,ji}(S_{ik},S_{kj},S_{ji})}}\rangle\\	
  	\end{aligned}
  	\end{equation}
  
  	where:
  	
  	\begin{widetext}
  	\begin{equation}\label{eq:A5}
  	\begin{aligned}	
  	&e^{-\beta H_{ik,kj,ji}(S_{ik},S_{kj},S_{ji})}=\\&
  	e^{-\beta H_{ik,kj,ji} (S_{ik},S_{kj},S_{ji}=1)}
  	-e^{-\beta H_{ik,kj,ji}(S_{ik}=1,S_{kj}=-1,S_{ji}=1)}
  	-e^{-\beta H_{ik,kj,ji} (S_{ik},S_{kj}=1,S_{ji}=-1)}
  	-e^{-\beta H_{ik,kj,ji}(S_{ik}=-1,S_{kj},S_{ji}=1)}+\\&e^{-\beta H_{ik,kj,ji}(S_{ik}=1,S_{kj},S_{ji}=-1)}
  	-e^{-\beta H_{ik,kj,ji} (S_{ik}=-1,S_{kj}=1,S_{ji}=-1)}
  	+e^{-\beta H_{ik,kj,ji}(S_{ik},S_{kj}=-1,S_{ji}=1)}
  	-e^{-\beta H_{ik,kj,ji}(S_{ik},S_{kj},S_{ji}=-1)}
  	\end{aligned}
  	\end{equation}	
  		\end{widetext}
  at least:\\
  	
  	\begin{widetext}
  		\begin{equation}\label{eq:A6}
  		\begin{aligned}
  		\langle S_{ik}S_{kj}S_{ji} \rangle& =\\&
  		 \left[\frac{
  			e^{\beta(N-3)(3q)+\beta}-3e^{\beta(N-3)(q)- \beta}{+3e^{\beta(N-3)(-q)+\beta}-e^{\beta(N-3)(-3q)-\beta}}}{{
  				e^{\beta(N-3)(3q)+\beta }+3e^{\beta(N-3)(q)-\beta}{+3e^{\beta(N-3)(-q)+\beta}+e^{\beta(N-3)(-3q)-\beta}}}}\right]
  		\end{aligned}
  		\end{equation}
  		\end{widetext}

  \end{appendices}
	
\end{document}